# Two Dimensional Allotropes of Arsenene with Wide Range of High and Anisotropic Carrier Mobility

Pooja Jamdagni [1*], Anil Thakur[2], Ashok Kumar[3], P.K.Ahluwalia[1*] and Ravindra Pandey[4]

[1]*Department of Physics, Himachal Pradesh University, Shimla, H.P. India, 171005*

[2]*Department of Physics, Govt. P. G. College, Solan, H.P. India, 173212*

[3] *Department of Physical Sciences, School of Basic and Applied Sciences, Central University of Punjab, Bathinda, India, 151001*

[4] *Department of Physics, Michigan Technological University, Houghton, MI, 49931, USA, 49931*

(November 13, 2018)

Emails:

j.poojaa1228@gmail.com (Pooja Jamdagni)

pk_ahluwalia7@yahoo.com (P. K. Ahluwalia)

**Abstract**

Considering the rapid development of experimental techniques for fabricating 2D materials in recent years, various monolayers are expected to be experimentally realized in the near future. Motivated by the recent research activities focused on the honeycomb arsenene monolayers, stability and carrier mobility of non-honeycomb and porous allotropic arsenene are determined using first principles calculations. In addition to five honeycomb structures of arsenene, a total of eight other structures are considered in this study. An extensive analysis comprising energetics, phonon spectra and mechanical properties confirms that these structures are energetically and dynamically stable. All these structures are semiconductors with a broad range of band gap varying from ~1 eV to ~2.5 eV. Significantly, these monolayer allotropes possess anisotropic carrier mobilities as high as several hundred $cm^2V^{-1}s^{-1}$ which is comparable with the well-known 2D materials such as black phosphorene and monolayer $MoS_2$. Combining such broad band gaps and superior carrier mobilities, these monolayer allotropes can be promising candidates for the superior performance of the next generation nanoscale devices. We further explore these monolayer allotropes for photocatalytic water splitting and find that arsenene monolayers have potential for usage as visible light driven photocatalytic water splitting.

# 1. Introduction

Due to variety of structures and fascinating electronic properties, two dimensional (2D) materials offer a promising scenario for the next generation electronic devices at nanoscale. After the successful experimental isolation of black phosphorene [1-2], and its exciting properties [3-6] have resulted into a growing interest in the exploration of novel allotropes of group-V monolayers [7-8]. Stability of the bulk phosphorus in variety of allotropic forms such as violet-, red-, white- and black-phosphorous have led to theoretical investigations of several allotropes of phosphorene [9-19], apart from the honeycomb α-P and β-P. All the polymorphs of phosphorene are predicted to be semiconductor with band gap ranging from 0.4 eV to 2.1 eV [10-19].

Extensive studies on phosphorene allotropes have encouraged the exploration of the next isoelectronic group-V elemental monolayer i.e. arsenene. Similar to α- and β-phosphorene, Kamal et. al. [20] predicted stable honeycomb structures of As, and Zhang et. al. [21] investigated the properties of β-phase of As. In addition, Mardanya et. al [22] also confirmed the stability of α-As and β-As with additional stable allotrope γ-As. Ma et. al. [23] proposed stable tricycle-type monolayer As, and Ersan et. al. [16] reported a stable square-octagon ring structure monolayer of group-V elements. Apart from the dynamical stability, α-and β-arsenene also exhibits high carrier mobility [24-26] with excellent thermoelectric and device performance [27-30]. The optical response of mono- and bi-layer arsenene (α- and β-phase) shows tunability with mechanical strains [31]. Few other atomically thin allotropes of arsenene were also reported to be dynamically stable [32-33].

In our previous study [34], we have identified the role of topological defects in the tunneling characteristics of group-V monolayers including arsenene. It is well known that the theoretical predictions often precede experimental synthesis and characterization owing to the difficulties of measurements at nanoscale. For example, predictions based on first principles calculations [1], have led to realization of the field effect transistor device consisting of a few layer thick black phosphorene [2]. Likewise, synthesis of blue phosphorene (β-P) [35] has been achieved after its prediction of stability by first principles methods [9, 36]. The rapid development in monolayer synthesis techniques are likely to pave the way for the fabrication of As-based monolayer allotropes in near future.

In this paper, we consider non-honeycomb (i.e. ε-, ξ- and 4-6-10-As) and porous (i.e. η-, θ-, hexstar- and K-As) allotropes of arsenene besides the widely studied α-, β-, γ-, δ-, tricycle-type- and square-octagon-As structures. The nomenclature of the arsenene allotropes was taken from that adopted for phosphorene allotropes. Density functional theory calculations were preformed to predict stability (both energetically and dynamically) and electronic properties of all the monolayer structures of arsenene, including the carrier transport properties using phonon-limited scattering model. Finally, as a case study from practical point of view, we have examined the possibility of application of these monolayers as water splitting photocatalysts.

## 2. Computational Model

Our theoretical analysis and predictions are based on density functional theory (DFT) as implemented in *Vienna ab-initio simulation package* (VASP) [37]. The electron exchange and correlation are treated within the framework of generalized gradient approximation (GGA) given by Perdew-Burke-Ernzerhof (PBE) functional [38]. We have used projector augmented-wave (PAW) method to describe electron-ion interaction [39]. Contributions from the van der Waals (vdW) interactions are also incorporated in calculations by using the DFT-D2 method of Grimme [40]. A plane wave basis set with kinetic energy cutoff of 400 eV was used. We adopt

Monkhorst-Pack scheme for k-point sampling of Brillouin zone integration with 12 x 12 x 1 mesh. In order to mimic the 2D system, we employ a supercell geometry with a vacuum of about 15 Å along z-direction (which is chosen perpendicular to the plane of arsenene). All the structures were fully relaxed using standard conjugate gradient method, with residual forces smaller than 0.001 eV/Å on each atom. The energy convergence value between two consecutive steps was chosen to be $10^{-6}$ eV. Since GGA-PBE functional underestimates the band gap, we have also applied a correction to PBE band structure using the screened hybrid functional HSE06 method [41-42].

The phonon-limited scattering model including the anisotropic characteristics of effective mass, elastic modulus and deformation potential was applied to calculate carrier mobilities of arsenene allotropes[24,30,43-44]:

$$\mu_{2D} = \frac{e\hbar^3 C^{2D}}{k_B T m^* m_a^* E_d^{i^2}} \qquad (1)$$

where $k_B$ is Boltzmann constant, $T=300\ K$ is temperature, $m^*$ is effective mass in transport direction (either $m^*_x$ or $m^*_y$ along the x and y direction, respectively) which is given as $m^* = \frac{\hbar^2}{\frac{\partial^2 E_0(k)}{\partial k^2}}$, where $E_0$ is the conduction/valance band energy along transport direction and $m_a^*$ is average effective mass determined by $m_a^* = \sqrt{m_x^* m_y^*}$. $E_d$ represents the deformation potential of hole/electron along transport direction determined by $E_d^i = \frac{\Delta E_i}{\Delta l/l_0}$ where $\Delta E_i$ is the energy change of the i$^{th}$ band under compression and dilation of cell. $C^{2D}$ is in-plane elastic modulus which can be derived from:

$$\frac{E-E_0}{S_0} = \frac{C^{2D}}{2}\left(\frac{\Delta l}{l_0}\right)^2 \qquad (2)$$

where $E - E_0$ represents the total energy change, $S_0$ is the area of xy-plane $\frac{\Delta l}{l_0}$ represents the deformation in x or y direction.

## 3. Results and Discussion

### 3.1 Structures

In Figure 1, the top and side views of monolayer allotropes of arsenene together with Wigner-Seitz cell are shown. We consider three types of monolayer geometric structures, namely: (i) honeycomb (ii) non-honeycomb and (iii) porous. The honeycomb structures are α-, β-, γ-, δ- and tricycle- type (T)-As, which contain the six membered rings are designated as (6-6)-As. T-As is constructed by the in-plane connections of the segments of α-As and β-As. The honeycomb structures of arsenene exhibit armchair and zigzag edges. Non-honeycomb structures are square-octagon (O)-As, ε-P and ξ-P which contain square and octagon units of arsenic atoms are denoted by (4-8)-As. For ε-As and ξ-As, each unit cell has two $As_4$ square units, and side views of both resemble the puckered armchair and zigzag structures of α-As and β-As, respectively. Similarly, (4-6-10)-As consists of four, six and ten member rings in non-honeycomb arrangements of atoms. Four porous allotropes namely η-As, θ-As, hex-star (HS)-As and K-As of arsenene are also shown in Figure 1. The As atoms in HS-As forms a hexagonal lattice with a Magen-David-like top view, whereas η-As and θ-As consists of the units of $As_5$ pentagons that are connected by interunit As-As bonds.

The structural and lattice parameters are listed in Table 1. The most of our calculated values are in very good agreement with the available values in literature, e.g., for α-As, our calculated lattice parameters (a=3.67 Å, b=4.77 Å), and thickness (h=2.40 Å) values agree well with the other reported values of lattice parameters (a=3.68 Å, b=4.77 Å) and thickness (h=2.39 Å) [20].

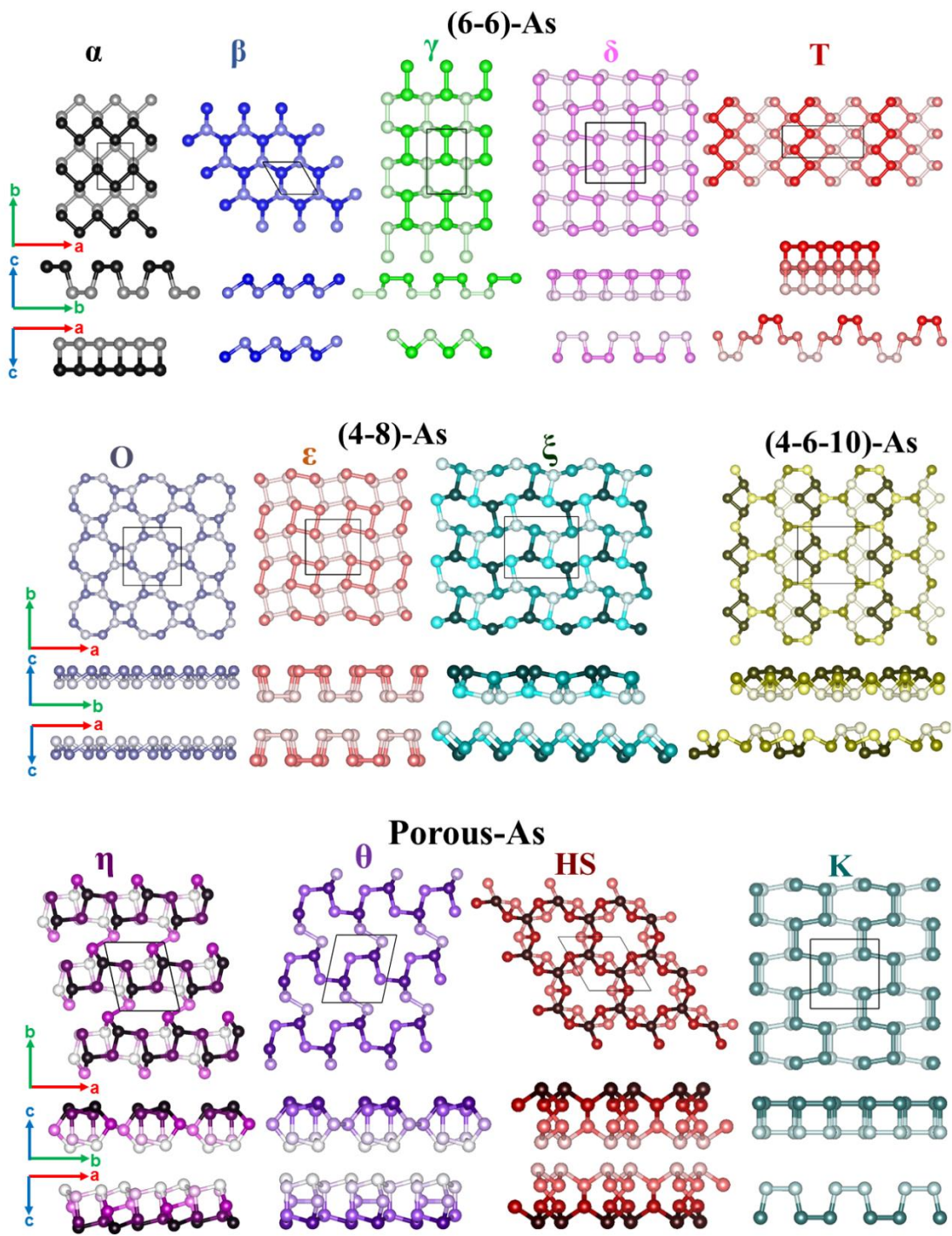

**Figure 1:** Top and side views of As-monolayer allotropes with five typical honeycomb 6-member ring structures (α, β, γ, δ, T); three non-honeycomb (4-8)-member rings structures (O, ε, ξ); one (4-6-10)-member rings structure and four porous structures (η, θ, HS, K). Atoms at the top and bottom of nonplanar layers are distinguished by color and shading of atoms (darkest color for the topmost atoms and lightest color for bottommost atoms). Two dimensional unit cell is also shown in the top view of each structure.

**Table 1:** Calculated values of equilibrium lattice parameters (a,b), thickness (h), bond lengths (d), bond angles (Θ), and cohesive energy ($E_c$) of arsenene allotropes using GGA-PBE functional including Waals (vdW) interaction term. *N* is number of atoms in an unit cell. Cohesive energy is calculated as: $E_c = |E_T - E_a|$, where $E_T$ and $E_a$ are total energy per atom of arsenene allotropes and total energy of isolated As atom. Other reported GGA-PBE values available in literature are also given.

| Phase | | Space group | N | Structural Parameters | | | | | $E_c$ eV/atom |
|---|---|---|---|---|---|---|---|---|---|
| | | | | a (Å) | b (Å) | h (Å) | $d_i$ (Å) | $\Theta_i$ (deg) | |
| (6-6)-As | α | Pmna | 4 | 3.67, 3.68[e], 3.68[a], 3.68[b] | 4.71, 4.77[e], 4.77[a], 4.82[b] | 2.39, 2.40[e], 2.39[b] | 2.48, 2.50, 2.49[a], 2.50[a], 2.50[b] | 94.6, 100.4, 94.6[a], 100.8[a], 94.5[b], 100.2[b] | 3.19, 3.13[b] |
| | β | P-3m1 | 2 | a=b=3.58, 3.60[e], 3.61[a], 3.61[b] | | 1.40, 1.38[e], 1.39[b] | 2.51, 2.50[a], 2.50[b] | 91.9, 92.2[a], 91.7[b] | 3.18, 3.15[b] |
| | γ | Pmmn | 4 | 3.57, 3.59[e], 3.58[b] | 5.88, 5.90[e], 5.92[b] | 1.69, 1.68[e], 1.68[b] | 2.49, 2.56, 2.50[b], 2.57[b] | 91.9, 98.8, 91.7[b], 98.7[b] | 3.13, 3.07[b] |
| | δ | Pbcm | 8 | 5.85, 5.90[e], 5.91[b] | 5.90, 5.94[e], 5.93[b] | 2.40, 2.38[e], 2.40[b] | 2.48, 2.52, 2.50[b] | 98.6, 100.0, 99.1[b], 99.8[b] | 3.15, 3.06[b] |
| | T | Pbcm | 8 | 9.36, 9.57[e], 9.58[c] | 3.65, 3.64[e], 3.65[c] | 4.41, 4.33[e], 4.34[c] | 2.49, 2.50, 2.51, 2.50[c], ,2.51[c] | 90.6, 93.5, 100.4, 90.5[c], 93.3[c], 101.41[c] | 3.20 |
| (4-8)-As | O | P4/nbm | 8 | a=b=7.06, 7.06[d] | | 1.41, 1.42[d] | 2.48, 2.52, 2.48[d], 2.52[d] | 71.7, 99.3, 71.8[d], 99.4[d] | 3.04, 2.98[d] |
| | ε | P4212 | 8 | a=b=5.84 | | 2.39 | 2.46, 2.56 | 90.0, 97.4, 101.5 | 3.10 |
| | ξ | P-1 | 8 | 7.05 | 5.86 | 1.69 | 2.49, 2.56 | 91.8, 98.9 | 3.11 |
| (4-6-10)-As | | Pmg | 10 | 8.45 | 6.67 | 2.45 | 2.49-2.52 | 81.6-100.9 | 3.09 |
| Porous-As | η | P-1 | 10 | 5.89 | 6.90 | 3.97 | 2.45-2.50 | 99.3-109.1 | 3.16 |
| | θ | Pm | 10 | 5.95 | 6.86 | 3.83 | 2.46-2.52 | 83.8-110.3 | 3.17 |
| | HS | P-31m | 10 | a=b=6.71 | | 4.45 | 2.50, 2.51 | 48.0-101.6 | 3.13 |
| | K | I2₁3 | 8 | 5.87 | 5.95 | 2.44 | 2.49, 2.50, 2.55 | 90.0, 99.8, 101.0 | 3.08 |

[a]ref. [20], [b]ref. [22], [c]ref. [23], [d]ref. [16], [e]our GGA-PBE calculated values

Notably, the shape diversities [Figure 1] of hexagonal rings in (6-6)-As and octagon rings is (4-8)-As originates from the bond angle variation while the bond lengths are found to be nearly same [Table 1]. The bond angle variation in (6-6)-As is as large as 20 degree which is more pronounced (~ 30 degree) in (4-8)-As structures. The bond length variations in both types of structures are calculated to be < 0.1 Å.

It is important to note that lattice constants 'b' in α-As and 'a' in δ- and T-As shows difference ≥ 0.05 Ấ in the calculated values using GGA+PBE and GGA+PBE+vdW level of theory. The thickness, h, in T-As is calculate to be 4.33 Ấ and 4.41 Ấ using GGA+PBE and GG+PBE+vdW method, respectively. The results therefore suggest that vdW correction is important to describe these layered structures with higher thickness. Throughout the study, we have used PBE functional with vdW correction and the overall results of the study remains nearly the same.

**3.2 Stability**

*a) Energetics*

In order to study the energetics of arsenene monolayers, we calculate their cohesive energy which is given in Table 1. Cohesive energy ($E_c$) is obtained as the difference between the total energy per atom of arsenene allotropes and energy of free As atom. Our calculated value of cohesive energy (3.19 eV/atom) for α-As is consistent with the previously reported value of 3.13 eV/atom [22]. Despite noticeable structural diversity in (6-6)-As and (4-8)-As allotropes, the cohesive energies are found to lie within a narrow range of 0.07 eV/atom. Among the various variants of arsenene, tricycle-type (T)-As possess the highest calculated value of cohesive energy (3.20 eV/atom), consequently it is the most stable allotrope followed by α-As, β-As, θ-As, η-As and δ-As. Note that the difference between the cohesive energies of α-, β- and T-As is very small (0.01-0.02 eV) [Table 1]. This is unlike phosphorous allotropes among which α-phase is found to be most stable [9, 36]. Next, we have calculated the formation energy relative to α-As using the formula:

$$RFE = \frac{E_A - NE_\alpha}{N} \qquad (3)$$

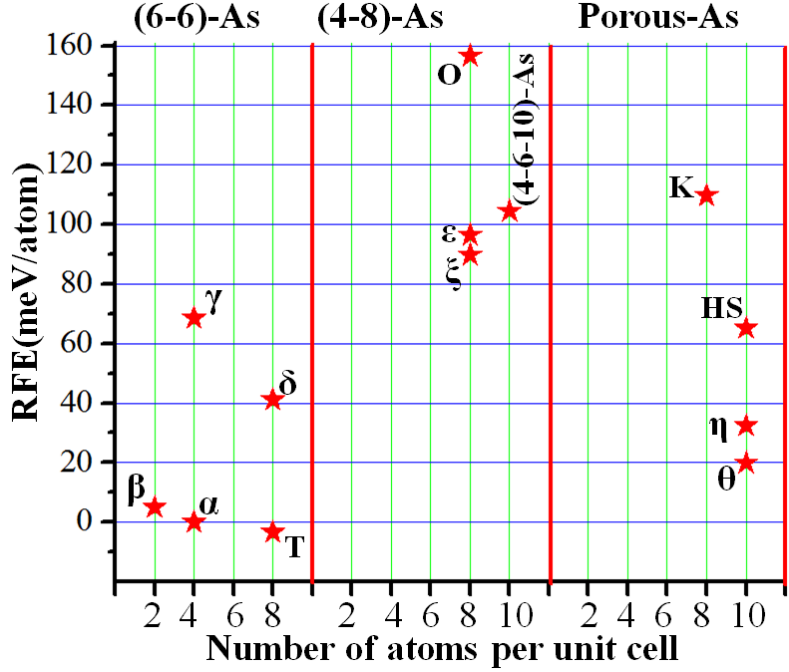

**Figure 2:** Relative formation energy (RFE) with respect to α-As, for various monolayer allotropes of As.

Here $E_A$ is the total energy of given monolayer allotrope, $E_\alpha$ is energy per atom of α-As monolayer and N is number of atoms per unit cell of given monolayer allotropes.

All the monolayer allotropes except O-As, (4-6-10)-As and K-As; have RFE less than 100 meV [Figure 2], indicating their ease of formation and energetic stability. T-As honeycomb monolayer structure has formation energy -3.41 meV/atom indicating its stability higher than α-As. Note that T-As is the in-plane heterostructure allotrope consisting of α and β allotropes of As. It is to be noted that the difference in the RFE of β-As, T-As and θ-As is less than 25 meV/atom [Figure 2], which is comparable to the thermal energy at room temperature. Note that ξ-As, ε-As, O-As, (4-6-10)-As and K-As possess high RFE ( > 80 meV) which is attributed to the increased tension in their structures due to 4-mebered rings. All other structures possess either five or six membered rings which makes them energetically more favorable. Looking at the existence of so many structures within very narrow range of RFE, it appears that phase coexistence may indeed be reality in the case of 2D allotropes of arsenene. Note that the bulk form of arsenic i.e. grey arsenic exists [45] and β-As monolayers can possibly be exfoliated from bulk As.

*b) Phonon Spectra*

Next, we computed the phonon dispersion spectra for arsenene allotropes and the results of phonon dispersion along the high symmetry points in the Brillouin zone are presented in Figure 3. From the phonon spectrum, it is possible to investigate dynamical stability of a given monolayer. The phonon frequencies are found to be positive for α-, β-, γ-, δ-, T- and ε-As, indicating their dynamical stability. Note that α-, β-, γ- and T-As are also found to be dynamically stable in the previous reported first principles calculations [20, 22-23]. In cases of O-, 4-6-10-, ξ-, η- and θ-As, all modes contain positive frequencies except out-of-plane acoustical modes near S and Y high symmetry points [Figure 3] due to the softening of phonons. Note that a similar prediction was made for β-As [20] and germanene [46]. On the other hand, HS- and K-As possess the out-of-plane acoustical modes with imaginary frequency in a large region of Brillouin zone [Figure 3], which indicates that these monolayers can only be stabilized on a suitable substrate which can damp the out-of-plane vibrations.

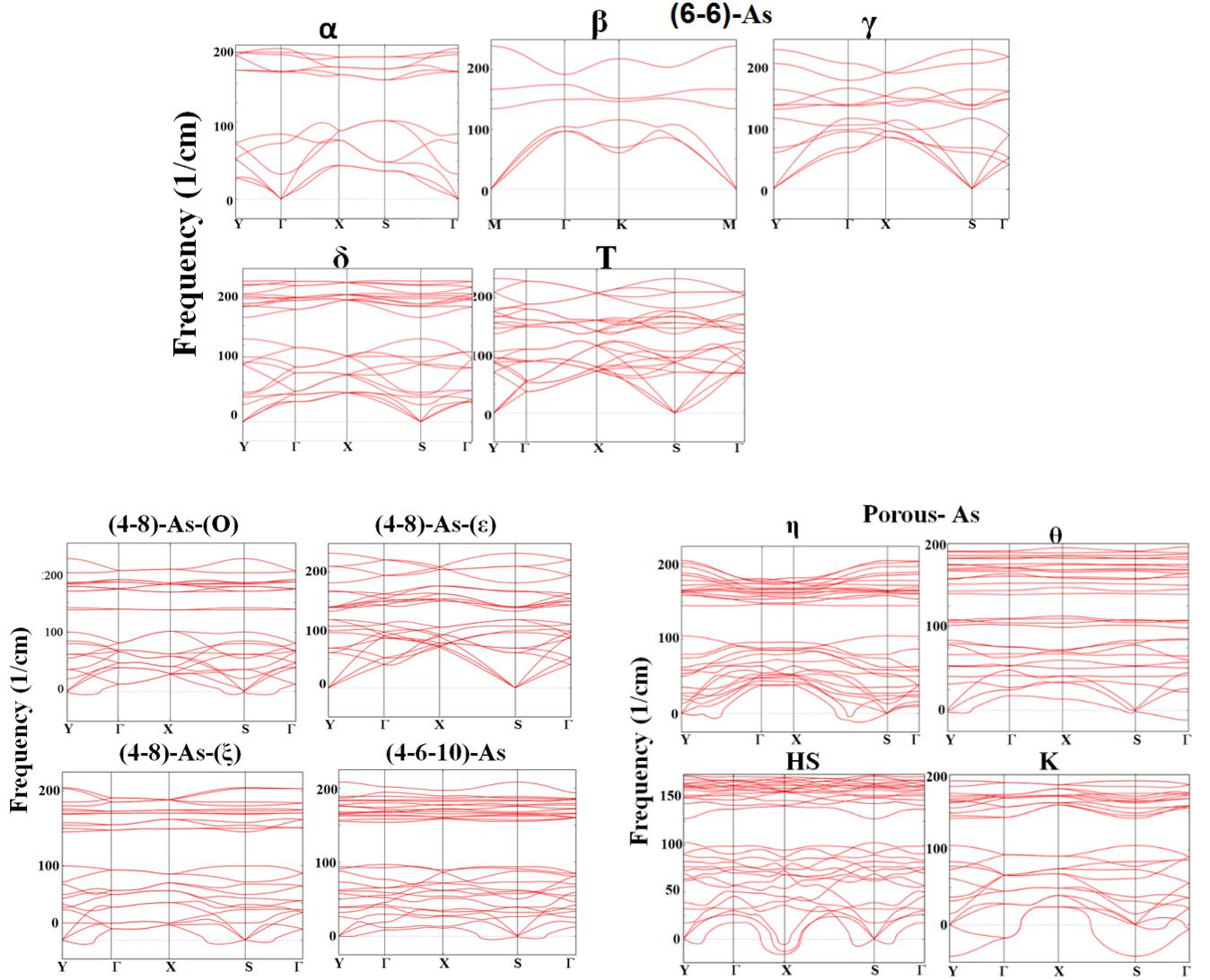

**Figure 3:** Phonon dispersion spectra for various monolayer allotropes of As.

*c) Mechanical Properties*

In-plane stiffness of a material is an important parameter to describe its mechanical stability. In order to describe mechanical properties of given monolayers, we calculate the elastic modulus by fitting strain energy density curve with in-plan strains [Figure 4]. Strain energy $(E - E_0)$ is obtained by applying a strain ($\pm$ 1.5%) in small steps of 0.05. Different strain energy density curves for strains along x- and y-direction are obtained due to the structural anisotropy of arsenene allotropes. For example, $C^{2D}_x$ and $C^{2D}_y$ for α-As are calculated to be 63.8 J/m$^2$ and 13.0 J/m$^2$, respectively, which are in very good agreement with the previously reported values of 65.2 J/m$^2$ and 16.9 J/m$^2$ [30]. β-As possess isotropic structure with a=b, our calculated value of elastic modulus is 55.5 J/m$^2$, which is in good agreement between the reported value of 51.4 J/m$^2$ [16]. The values of elastic modulus are listed in Table 2. Anisotropic elastic modulus is also evident from curvature of strain-energy density versus strain curve [Figure 4], e.g., large curvature of strain along x-direction in η-As leads to higher value of elastic modulus (82.7 J/m$^2$) as compared to small curvature of strain along y-direction which gives rise to smaller value of elastic modulus (40.8 J/m$^2$). The calculated values of the elastic modulus are much smaller than those of some typical 2D materials such as graphene (330 J/m$^2$) [7], h-BN (240 J/m$^2$) [47] and MoS$_2$ (180 J/m$^2$) [48], suggesting relatively superior flexibility of arsenene monolayers.

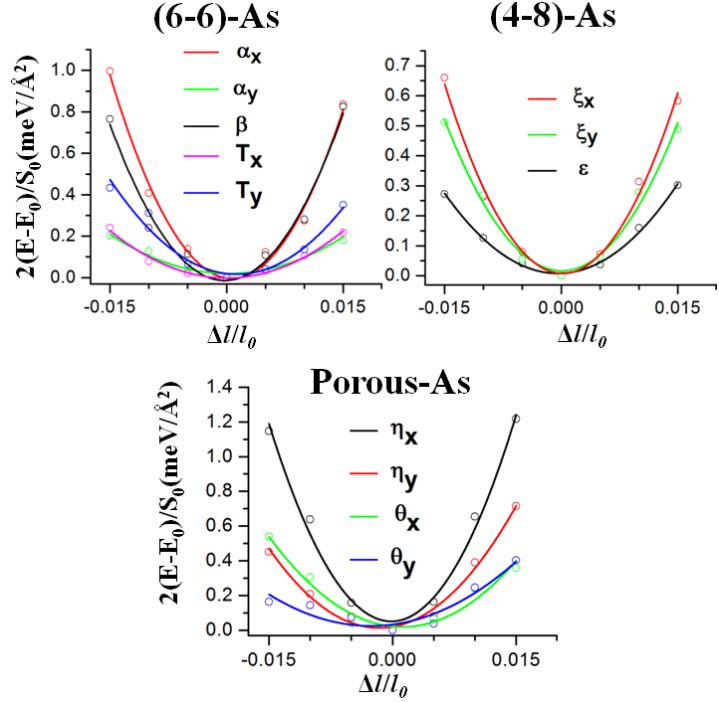

**Figure 4:** The 2D elastic modulus obtained from fitting strain energy density curve $\frac{2(E-E_0)}{S_0}$ versus $\Delta l/l$ for low energy allotropes of monolayer As. Uniaxial strain is applied along x and/or y direction.

**Table 2:** Calculated values of elastic modulus ($C^{2D}$) along x and y-direction. Band gap ($E_g$) and workfunction (φ) using the HSE functional for monolayer allotropes of As.. Previously reported values are also listed.

| Structure | | $C^{2D}$ (J/m²) | | $E_g$-(HSE)(eV) | Φ (HSE) (eV) |
|---|---|---|---|---|---|
| | | x | y | | |
| (6-6)-As | α | 63.8, 65.2[a] | 13.0, 16.9[a] | 1.38 | 4.77 |
| | β | 55.5, 51.4[b] | | 2.21 | 5.55 |
| | γ | 49.7 | 27.6 | 1.42 | 5.18 |
| | δ | 7.7 | 6.4 | 1.09 | 4.74 |
| | T | 16.2 | 27.5 | 1.14 | 4.65 |
| (4-8)-As | O | 11.4, 20.9[b] | | 2.48, 2.47[b] | 5.75 |
| | ε | 35.4 | | 1.36 | 5.11 |
| | ξ | 43.8 | 20.1 | 1.43 | 5.30 |
| (4-6-10)-As | | 6.0 | 6.8 | 1.63 | 5.32 |
| Porous-As | η | 82.7 | 40.8 | 1.64 | 5.09 |
| | θ | 31.9 | 19.0 | 1.77 | 5.17 |
| | HS | 3.0 | | 1.74 | 5.22 |
| | K | 2.5 | 1.4 | 1.07 | 4.67 |

[a]ref. [30], [b]ref. [16]

### 3.3 Electronic Band Structures

Allotropes of monolayer arsenic are predicted to be semiconductors with a broad range of band gaps which are important for broadband photo-response. γ-As, T-As, O-As and ε-As are direct gap semiconductors while the other monolayers are indirect gap semiconductors [Figure 5]. Our calculated values of band gaps are in very good agreement with the available values in literature, e.g, indirect band gap of α-As (0.72 eV) and β-As (1.58 eV) are in very good agreement with the previously reported values of 0.77 eV and 1.57 eV, respectively [16]; direct band gap calculated for γ-As (0.79 eV) and O-As (1.81 eV) are consistent with the other reported values 0.86 eV [22] and 1.79 eV [16], respectively. The calculated values vary from 0.37 eV in K-As to 1.81 eV in O-As. Additionally, the HSE06 hybrid functional calculations predict significantly larger values of band gaps [Table 2] ranging from ~ 1 to ~2.5 eV without changing the nature of the band gaps.

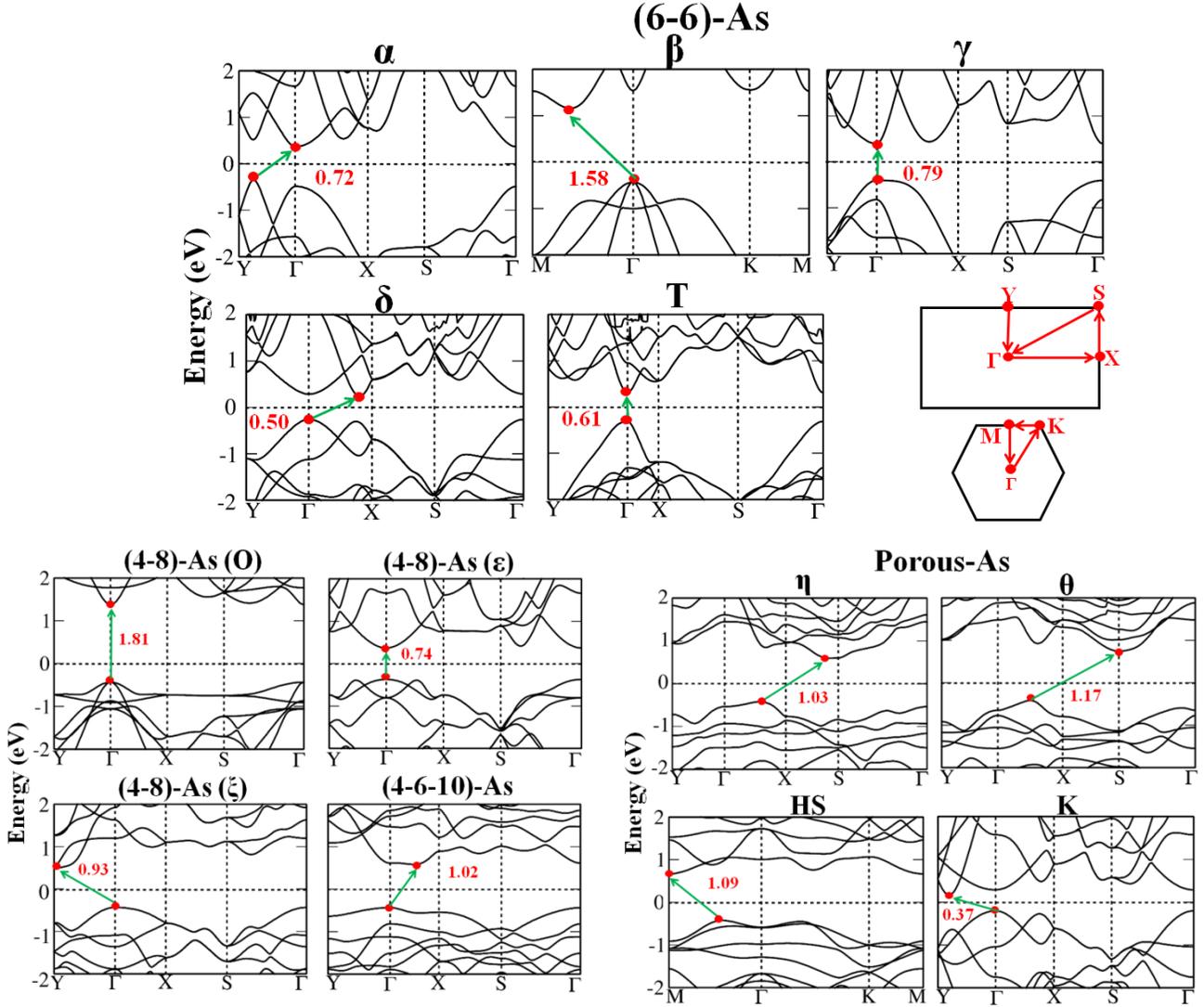

**Figure 5:** Electronic band structure of various monolayer allotropes of As calculated using GGA-PBE method. Fermi level is set to zero. The Brillouin zones with high symmetry directions are also shown. β-As and HS-As possess hexagonal Brillouin zone while rest of monolayers have rectangular Brillouin zone.

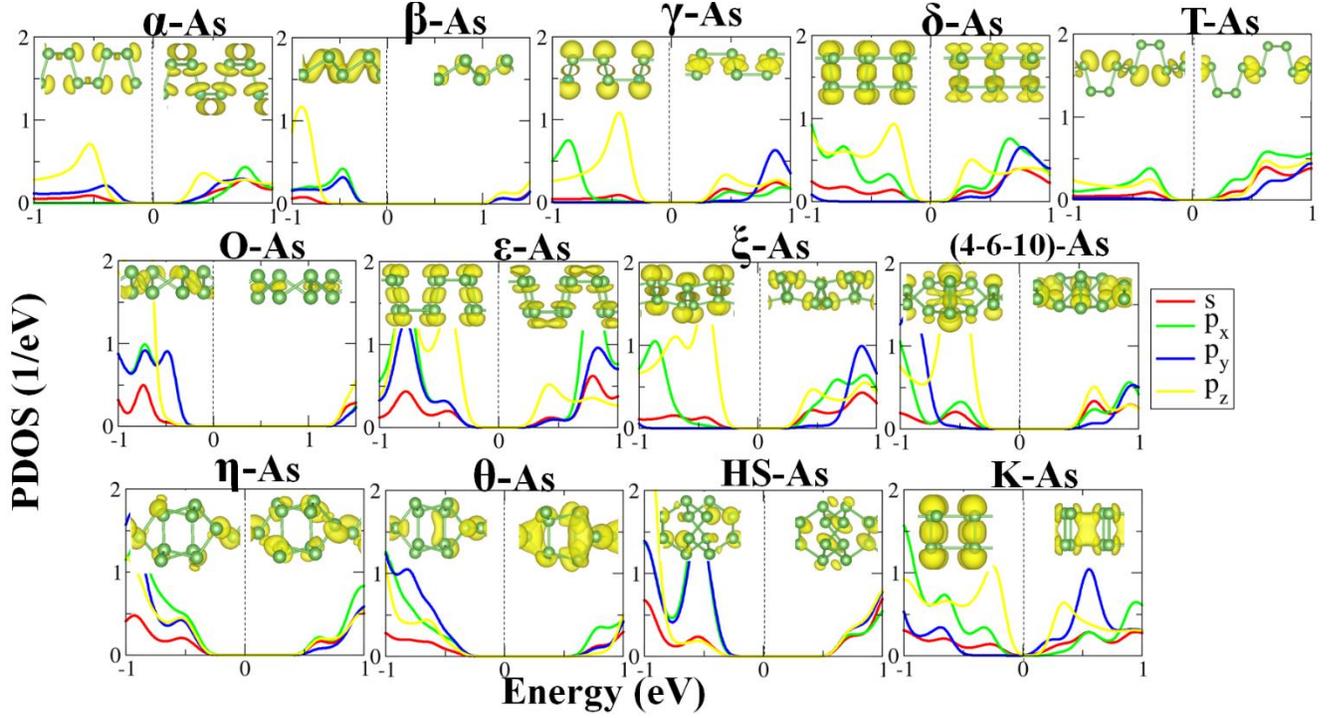

**Figure 6:** The orbital projected density of states (PDOS) for the various allotropes of arsenene calculated using GGA-PBE functional. The p orbitals are dominant for all structures. The vertical dotted line in each plot denote Fermi energy which is set at 0 eV. The VBM and CBM charge densities distributions (side views) are represented at the isosurface of $3\times10^{-4}$ e/Å$^3$ and superposed with a ball-and-stick model of the structure. Charge density profile for valance and conduction band edge are located below and above the Fermi level, respectively, in PDOS plots.

In order to understand the contribution of different orbitals to the electronic states near Fermi level, we calculate the orbitals projected density of states (PDOS), which reveals that the states near the valance and conduction-band edges are primarily contributed by p orbitals [Figure 6]. The fact that the p orbitals are dominant is a common feature of monolayer honeycomb systems such as silicene, germanene and phosphorene, where sp$^2$-like bonds form non-planar honeycomb structure. Notable contribution from p$_z$ orbitals with out-of-plan electron density perpendicular to monolayer can be seen in PDOS plots. This is further endorsed by the special distribution of VBM and CBM charge density and can be characterized by the localization of charge on the atomic sites [Figure 6]. In the case of multilayer structures, these states are going to participate in interlayer hopping with adjacent layers, resulting in a band dispersion perpendicular to the plane. Therefore, multilayer structures of these allotropes are expected to have smaller band gap than their monolayer structures. Similar effect is observed in phosphorous [36] and transition metal dichalcogenides monolayers [49], where the band gap is found to be inversely proportional to the number of layers.

A comparison between arsenene allotropes and their predecessor phosphorene allotropes reveals some common features in the electronic band structures due to the similarity in their crystal structures and electronic configurations. For example, energy dispersions of the valance and conduction bands are comparable for similar allotropes belonging to arsenene and phosphorene [9, 11-12, 16-17]. As a results, $\beta$-, $\xi$-, $\eta$-, $\theta$- and HS-As allotropes of arsenene and phosphorene are qualitatively similar being indirect band gap semiconductors. Although the valance and conduction band energies are very close to the respective maximum and minimum at multiple points in Brillouin zones for both arsenene and phosphorere, the valance band maximum (VBM) and conduction band minimum (CBM) are located at different points for equivalent arsenene and phosphorene allotropes, e.g., location of VBM in $\beta$-As is at $\Gamma$ whereas for $\beta$-P it is in between $\Gamma$-K direction; location of CBM in $\xi$-As and $\xi$-P is at Y and X, respectively. This difference is attributed to the more delocalization of valance band edge charge density at $\Gamma$ point for $\beta$-As and conduction band edge charge density at Y point for $\xi$-As, that leads to the shift in the band energy towards Fermi level. The more delocalization at these band edges are due to the hybridization between the in-plane and out-of-plane p orbitals that spatially extend the charge distribution away from the atomic sites [Figure 6]. Both ε-As and ε-P are direct gap semiconductor with direct transition at Γ point. Unlike phosphorene, α- and δ-As monolayers are indirect gap semiconductors, whereas γ-, T- and O-As are direct gap semiconductors. This difference in the nature of band gap of two systems may also be attributed to the weaker electronegativity of As atom as compared to P atom, that leads to more delocalization of charge density in As systems and hence decrease in VBM/CBM energy towards Fermi level.   It is important to note that the difference between direct and indirect band gap in γ-, δ-, O-, ε- and (4-6-10)-As is very small (< 0.03 eV), and a small value of the in-plane mechanical strain can induce the direct↔indirect gap transition [50] in these monolayers.

**3.4 Carrier Mobilities**

The carrier mobility is a property that is directly related with the electronic conductivity of 2D materials. In the framework of the longitudinal-acoustic phonon dominated scattering mechanism, the carrier mobilities ($\mu_{2D}$) can be calculated using values of the elastic modulus, effective masses and deformation potentials [Equation 1]. $\mu_{2D}$ is directly proportional to elastic modulus, that is found to be highly anisotropic [Table 2] for all the monolayer allotropes except β-As, O-As, ε-As and HS-As. Essentially, the structural anisotropy in arsenene allotropes leads to the anisotropic carrier transport, thereby suggesting that a highly flexible monolayer can give rise to less carrier mobility and vice versa. And, the anisotropic elastic modulus can be attributed to the bonding structure of these allotropes. Under uniaxial strain along y-direction, the bond angles can be modified than those along x-direction that may lead to different superposition of atomic orbitals in both x- and y- directions.  Furthermore, the carrier mobility not only depends on the mechanics of monolayer, but has also strong dependence on the carrier effective masses. Note that $\mu_{2D}$ is inversely proportional to the product of the carrier effective masses along transport direction and the average effective masses. The carrier effective masses along x- and y-direction are calculated by fitting $E$ vs. $k$ diagram along Γ-X and Γ-Y direction, respectively and are listed in Table 3. Our calculated values of electron and hole effective masses for α-As ($m_{e-x}^*$=1.29, $m_{e-y}^*$= 0.39, $m_{h-x}^*$=1.57 and $m_{h-y}^*$= 0.21) are in close agreement with the previously reported values ($m_{e-x}^*$=1.17, $m_{e-y}^*$=0.24, $m_{h-x}^*$=1.64, $m_{h-y}^*$=0.13) [30]. For β-As, $m_e^*$ and $m_h^*$ is calculated to be 0.13 and 0.53, respectively, which is in excellent agreements with the other reported values 0.128 and 0.501 [27]. Due to the highly isotropic structural arrangement of As atoms in the honeycomb lattice of β-As, it has nearly the same carrier effective masses along x- and y-directions which is also evident from the symmetric $E$ vs. $k$ dispersion around Γ in the electronic band structure of β-As [Figure 5]. The carrier effective masses are calculated to be highly anisotropic for the monolayer allotropes having structural anisotropy. For example the effective masses of electron (holes) for ξ-As and η-As along Γ-X/Γ-Y direction are calculated to be 0.55/4.85 (1.73/1.52) and 1.40/3.47 (3.52/1.68). The carrier effective masses at VBM/CBM are listed in Table 3.

Table 3: Calculated carrier effective masses (m*) of electron and hole at band edges. The $m_e^*$ ($m_h^*$) represent electron (hole) effective masses, along Γ-X and Γ-Y direction of Brillouin zone. Deformation potential of electron ($E^e_d$) and hole ($E^h_d$) along x and y direction is given. The carrier effective masses and deformation potential for few monolayer allotropes have same value along x and y direction, so single value for these monolayers is given.

| Structure | | m*(m) | | $m_e^*$(m) | | $m_h^*$(m) | | $E^e_d$ (eV) | | $E^h_d$ (eV) | |
|---|---|---|---|---|---|---|---|---|---|---|---|
| | | CBM | VBM | x | Y | x | y | X | y | x | Y |
| (6-6)-As | A | 0.78 | 0.16 | 1.29, 1.17[a] | 0.39, 0.24[a] | 1.57, 1.64[a] | 0.21, 0.13[a] | 3.5 | 1.4 | 2.1 | 7.2 |
| | B | 0.81 | 0.52 | 0.13, 0.128[b] | | 0.53, 0.501[b] | | 3.9 | | 1.7 | |
| | Γ | 0.35 | 1.08 | 0.20 | 1.10 | 1.11 | 2.12 | 7.9 | 9.2 | 7.8 | 2.8 |
| | Δ | 0.67 | 0.54 | 0.96 | | 0.69 | | 3.2 | 4.8 | 2.3 | 3.8 |
| | T | 0.15 | 0.35 | 0.48 | 0.28 | 0.81 | 13.38 | 4.9 | 3.2 | 3.9 | 3.1 |
| (4-8)-As | O | 0.13 | 0.62 | 0.22 | | 4.70 | | 2.2 | | 0.2 | |
| | E | 0.67 | 0.80 | 0.91 | | 1.18 | | 1.7 | | 2.9 | |
| | Ξ | 0.45 | 1.09 | 0.55 | 4.85 | 1.73 | 1.52 | 1.9 | 6.2 | 0.2 | 0.2 |
| (4-6-10)-As | | 0.52 | 1.88 | 5.51 | 0.32 | 3.28 | 4.02 | 2.9 | 0.8 | 0.7 | 0.6 |
| Porous-As | H | 0.76 | 0.75 | 1.40 | 3.47 | 3.52 | 1.68 | 4.6 | 4.7 | 0.3 | 0.4 |
| | Θ | 0.19 | 0.48 | 0.74 | 25.98 | 25.81 | 0.68 | 5.7 | 5.6 | 0.2 | 0.3 |
| | HS | 21.3 | 0.49 | 0.88 | | 5.33 | | 0.7 | | 0.2 | |
| | K | 0.11 | 0.69 | 0.89 | 0.89 | 0.68 | 0.68 | 2.1 | 1.1 | 1.9 | 0.9 |

[a]ref. [30], [b]ref. [27]

Another key factor which also determines the carrier mobility is the deformation potential ($\mu_{2D} \propto 1/E_d^2$) and is calculated by the linear fitting of the VBM)/CBM vs. strain surface. The magnitude of $E_d$ describes the change in energy of the electronic band with the elastic deformation and, therefore, describes the degree to which the charge carriers interact with phonons. A lower value of $E_d$ indicates a weaker electron-phonon coupling in the conduction (valence) band, thereby contributing to an increase in the mobility of electrons (holes) [43-44]. The deformation potential shows significant dependence on both transport direction and types of carriers. For example for α-As, the value of $E_d$ for holes along y-direction is 3.5 times more than the value in x-direction whereas the value of $E_d$ for electrons along y-direction is 2.5 times less than the value in x-direction. The values of $E_d$ for electrons and holes along x- and y-directions are listed in Table 3.

Incorporating the anisotropic characteristics of elastic modulus, effective masses and deformation potentials, we have calculated the carrier mobilities at T=300 K. Our calculated values of carrier mobilities show directional anisotropy and are found to be, in general, moderately high [Figure 7]. For example hole (electron) mobility of α-As along x-direction is 7.5 (4) times higher than the mobility along y-direction. The hole (electron) mobilities of ξ-As varies from 4 x $10^3$ to 8 x $10^3$ (0.02 x $10^2$ to 3 x $10^2$) $cm^2V^{-1}s^{-1}$. The highest hole mobility of the order of $10^3$ $cm^2V^{-1}s^{-1}$ has been obtained for ξ-As, η-As, θ-As [Figure 7]. The high hole mobility of these monolayers are mainly due to their very small deformation potentials (0.2-0.7 eV). The small deformation potentials of these monolayer allotropes correspond to the localized VBM charge carrier density [Figure 6] that results into almost no change to the energy of these states by the displacement of phonons. Likewise, the highest electron mobility of the order of $10^3$ $cm^2V^{-1}s^{-1}$ is calculated for β-As and O-As which is attributed to their relatively small electron effective masses (0.13 for β-As and 0.22 for O-As) [Table 3]. In order to find out the extent of anisotropy in carrier mobilities, we now calculate the mobility anisotropy which is defined as:

$$R_a = \frac{Max(\mu_x, \mu_y)}{Min(\mu_x, \mu_y)} \qquad (4)$$

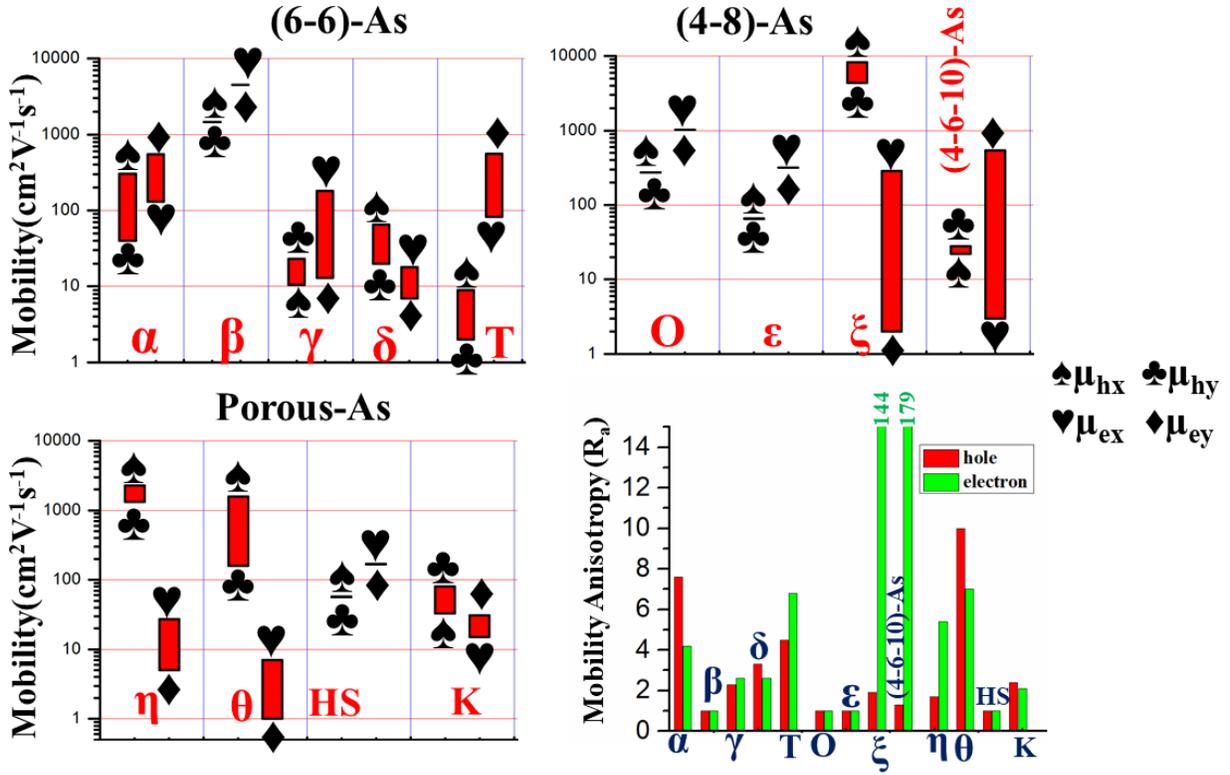

**Figure 7:** The room temperature carrier mobilities of monolayer allotropes of As. $\mu_{hx}$ and $\mu_{hy}$ represent hole mobility along x and y direction while $\mu_{ex}$ and $\mu_{ey}$ represent electron mobility along x and y direction. Each bar for electron or hole mobility represents the upper and lower limit of the mobility of electron or hole in a particular structure. Mobility anisotropy ($R_a$) is calculated as: $R_a$= Max ($\mu_x$, $\mu_y$)/Min ($\mu_x$, $\mu_y$).

$R_a$ is equal to 1.0 for isotropic systems and is larger than 1.0 for anisotropic systems. Mobility anisotropy for both hole and electron is given in Figure 7. $R_a$ is calculated to be 1.0 for β-As, O-As, ε-As and HS-As indicating isotropic carrier transport in these monolayers. Among the monolayers allotropes considered, mobility anisotropy of electrons in ξ-As (144) and (4-6-10)-As (179) is calculated to be of highest values. Mobility anisotropy of carriers for all other monolayer allotropes has been found between $1 < R_a < 10$.

It is worth comparing the calculated carrier mobilities of arsenene monolayers with important 2D materials such as graphene, black phosphorene and MoS$_2$. Graphene possess highest carrier mobility of the order of $10^4$ cm$^2$V$^{-1}$s$^{-1}$, whereas black phosphorene and monolayer MoS$_2$ exhibit carrier mobility of the order of $10^3$ cm$^2$V$^{-1}$s$^{-1}$ and $10^2$ cm$^2$V$^{-1}$s$^{-1}$, respectively [51]. We found that electron and hole mobility of β-As; hole mobility of ξ-As, η-As and θ-As; and electron mobility of O-As, are comparable with black phosphorene, whereas, electron and hole mobility of α-As; and electron mobility of γ-As, T-As, ε-As, (4-6-10)-As and HS-As are comparable with monolayer MoS$_2$. Note that the high intrinsic carrier mobilities plays crucial role in the performance of nanoelectronic devices based on layered structures [1-2].

**3.5 Potential Application in Photocatalytic Water Splitting**

We now investigate the feasibility of the monolayer allotropes of arsenic to be used for photocatalytic water splitting. In general, the band gap of more than 1.23 eV is required for the photocatalysis reaction. At ambient conditions, CBM must locate itself more negative than the redox potential of $H^+/H_2$, 0 V vs normal hydrogen electrode (NHE) and the VBM must align more positive than the redox potential $O_2/H_2O$ (1.23 V vs NHE) [52]. Band alignments of monolayer allotropes of arsenic with respect to vacuum level in neutral medium (pH=7) are presented in Figure 8. It is found that the band alignments and band gaps of β-As, γ-As, O-As, (4-6-10)-As, η-As, θ-As and HS-As allotropes matches well with the redox potential of water, therefore, these monolayers can potentially be used for photocatalytic water splitting. Furthermore, their band gaps except γ-As, lies in the visible region, so these 2D materials have a promise and novelty to harvest visible light for photocatalytic splitting of water. It has been shown previously that strain engineering can tune the band gaps and band alignments of monolayers [53], which further enhances the scope and usage of studied monolayer allotropes of arsenic for visible light driven photocatalytic water splitting. Despite known toxicity of arsenene, it has emerged as a promising material for the next-generation electronic and optoelectronic applications [54-55]. Recent fabrication of 2D arsenene nanosheets [56] is likely to pave the way for the use of arsenene in water splitting photocatalysis as suggested by this work.

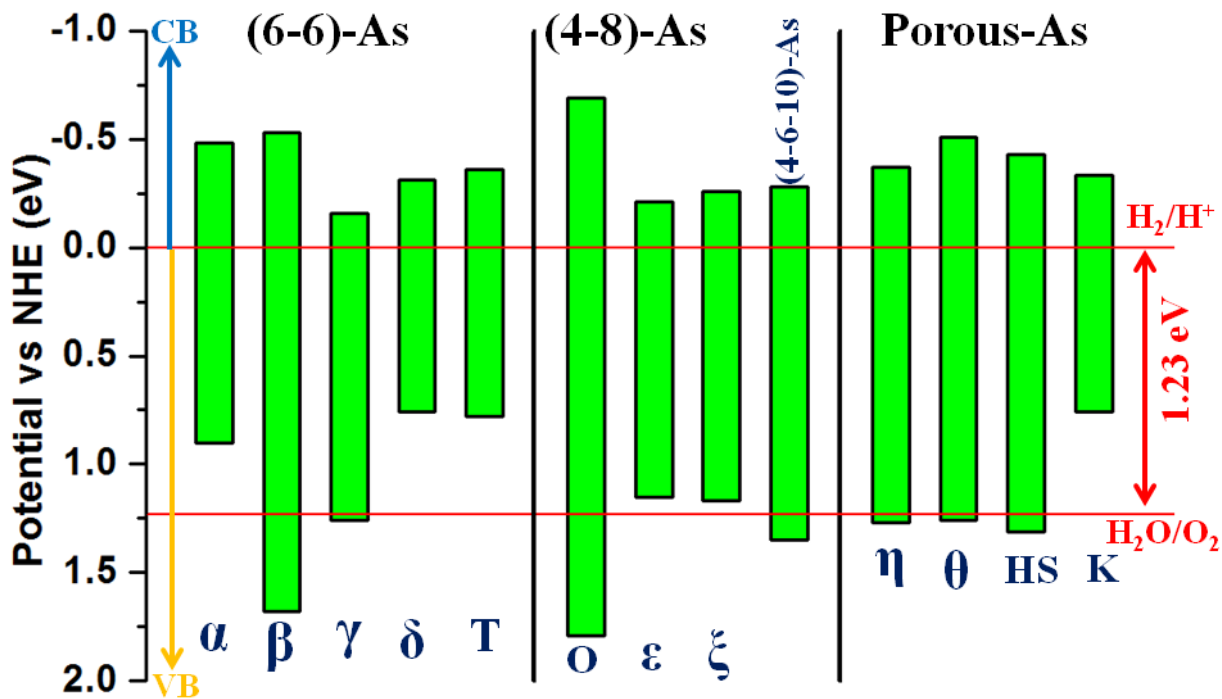

**Figure 8:** Band energy alignment of valance band maximum (VBM) and conduction band minimum (CBM) for monolayer allotropes of As at ambient conditions. The band edges relative to vacuum are calculated using the HSE06 functional. Horizontal red lines represent water redox potential.

## 4. Conclusions

We systematically investigated a wide variety of monolayer allotropes of arsenic using density functional theory. Through energetics and dynamical stability analysis we predict that besides honeycomb structures, the non-honeycomb and porous structures of monolayer arsenic are also stable. Results of cohesive energy and relative formation energy calculations show that, these monolayer allotropes have the possibility of experimental fabrication. These monolayer structures are also found to be mechanically flexible and structural anisotropy gives rise to anisotropic elastic properties. We find that the considered monolayer allotropes cover a wide range of band gaps, which can be helpful for broad band photo-response. More importantly, these monolayers possess anisotropic carrier mobilities as high as several hundred square centimeters per volt-second. The high intrinsic mobilities of these monolayer allotropes may play crucial role in the performance of nanoelectronic and optoelectronic devices based on arsenene. Considering the wide range of band gaps and high carrier mobility, these monolayers are found to be potential candidates for visible light driven photocatalytic water splitting.

## Acknowledgements

PJ is grateful to UGC New Delhi for providing financial assistance in the form of UGC-BSR senior research fellowship. Usage of CV Raman HPC cluster (DST, New Delhi, Govt. of India funded) at Himachal Pradesh University, Shimla and Superior High Performance Computing cluster at Michigan Technological University Houghton, USA is gratefully acknowledged for obtaining results presented in this paper

## References

1. H. Liu, A. T. Neal, Z. Zhu, Z. Luo, X. Xu, D. Tomanek and P. D. Ye, Phosphorene: An Unexplored 2D Semiconductor with a High Hole Mobility, *ACS Nano,* 2014 **8**, 4033-4041.

2. L. Li, Y. Yu, G. J. Ye, Q. Ge, X. Ou, H. Wu, D. Feng, X. H. Chen, Y. Zhang, Black phosphorus field-effect transistors, *Nat. Nanotechnol.,* 2014 **9**, 372-377.

3. J. Qiao, X. Kong, Z. X. Hu, F. Yang and W. Ji, High-mobility transport anisotropy and linear dichroism in few-layer black phosphorus, *Nat. Commun.,* 2014 **5**, 4475

4. J. W. Jiang and H. S. Park, Negative poisson's ratio in single-layer black phosphorus, *Nat. commun.* 2014, **5**, 4727.

5. Q. Liu, X. Zhang, L. B. Abdalla, A. Fazzio and A. Zunger, Switching a normal insulator into a topological insulator via electric field with application to phosphorene, *Nano Lett.*, 2015 15, 1222-1228.

6. R. Fei, A. Faghaninia, R. Soklaski, J. A. Yan, C. Lo and L. Yang, Enhanced thermoelectric efficiency via orthogonal electrical and thermal conductances in phosphorene, *Nano Lett.*, 2014, **14**, 6393-6399.

7. S. Zhang, S. Guo, Z. Chen, Y. Wang, H. Gao, J. G. Herrero, P. Ares, F. Zamora, Z. Zhu and H. Zeng, Recent progress in 2D group-VA semiconductors: from theory to experiment, *Chem. Soc. Rev.*, 2018 **47**, 982-1021.

8. M. Pumer and Z. Sofer, 2D Monoelemental Arsenene, Antimonene, and Bismuthene: Beyond Black Phosphorus, *Adv. Mater.,* 2017, **29**, 1605299.

9. J. Guan, Z. Zhu and D. Tomanek, Phase coexistence and metal insulator transition in few-layer phosphorene: A computational study, *Phys. Rev. Lett.*, 2014 **113** 046804.

10. H. Wang, L. Xingxing, L. Pai and J. Yang, δ-phosphorene: a two dimensional material with high negative Poisson's ratio, *Nanoscale*, 2017, **9**, 850-855.

11. M. Wu, H. Fu, L. Zhou, K. Yao and X. C. Zeng, Nine new phosphorene polymorphs with non-honeycomb structures: A much extended family, *Nano Lett.*, 2015, **15**, 3557.

12. T. Zhao, C. Y. He, S. Y. Ma, K. W. Zhang, X. Y. Peng, G. F. Xie and J. X. Zhong, A new phase of phosphorus: the missed tricycle type red phosphorene, *J. Phys.: Condens. Matter*, 2015, **27** 265301.

13. Y. Zhang, Z. Wu, P. Gao, D. Fang, E. Zhang and S. Zhang, Structural, elastic, electronic, and optical properties of the tricycle-like phosphorene, *Phys. Chem. Chem. Phys.*, 2017, **19**, 2245-2251.

14. H. Wang, X. Li, Z. Liua and J. Yanga, ψ-phosphorene: a new allotrope of phosphorene, *Phys. Chem. Chem. Phys.*, 2017, **19** 2402.

15. P. Li and W. Luo, A new structure of two-dimensional allotropes of group V elements, *Sci. Rep.*, 2016, **6** 25423.

16. F. Ersan, E. Akturk and S. Ciraci, Stable single-layer structure of group-V elements, *Phys. Rev. B*, 2016, **94**, 245417.

17. M. Xu, C. He, C. Zhang, C. Tang and J. Zhong, First-principles prediction of a novel hexagonal phosphorene allotrope, *Phys. Status Solidi RRL*, 2016, **10** 563.

18. Z. Zhuo, X. Wu and J. Yang, Two-Dimensional Phosphorus Porous Polymorphs with Tunable Band Gaps, *J. Am. Chem. Soc.*, 2016, **138** 7091-7098.

19. C. He, C. X. Zhang, C. Tang, T. Ouyang, J. Li and J. Zhong, Five low energy phosphorene allotropes constructed through gene segments recombination, *Sci. Rep.*, 2017, **7**, 46431.

20. C. Kamal and M. Ezawa, Arsenene: Two-dimensional buckled and puckered honeycomb arsenic systems, *Phys. Rev. B*, 2015, **91**, 085423.

21. S. Zhang, Z. Yan, Y. Li, Z. Chen, H. Zeng, Atomically thin arsenene and antimonene: semimetal-semiconductor and indirect-direct band-gap transitions, *Angew. Chem., Int. Ed.*, 2015, **54**, 3112-3115.

22. S. Mardanya, V. K. Thakur, S. Bhowmick and A. Agarwal, Four allotropes of semiconducting layered arsenic that switch into a topological insulator via an electric field: Computational study, *Phys. Rev. B*, 2016, **94**, 035423.

23. S. Y. Ma, P. Zhou, L. Z. Sun and K.W. Zhang, Two-dimensional tricycle arsenene with a direct band gap, *Phys. Chem. Chem. Phys.*, 2017, **18**, 8723-8729.

24. S. Zhang, M. Xie, F. Li, Z. Yan, Y. Li, E. Kan, W. Liu, Z. Chen and H. Zeng, Semiconducting group 15 monolayers: A broad range of band gaps and high carrier mobilities, *Angew. Chemie*, 2015, **127**, 1-5.

25. Z. Zhang, J. Xie, D. Yang, Y. Wang, M. Si and D. Xue, Manifestation of unexpected semiconducting properties in few-layer orthorhombic arsenene, *App. Phys. Exp.*, 2015, **8**, 055201.

26. C. Wang, Q. Xia, Y. Nie, M. Rahman and G. Guo, Strain engineering band gap, effective mass and anisotropic Dirac-like cone in monolayer arsenene, *AIP Adv.*, 2016, **6**, 035204.

27. G. Pizzi, M. Gibertini, E. Dib, N. Marzari, G. Iannaccone and G. Fiori, Performance of arsenene and antimonene double-gate MOSFETs from first principles, *Nat. Commun.*, 2016, **7**, 12585.

28. S. Sharma, S. Kumar, and U. Schwingenschlogl, Arsenene and Antimonene: Two-dimensional materials with high thermoelectric figures of Merit, *Phys. Rev. Appl.*, 2017, **8**, 044013.

29. Y. Wang, P. Huang, M. Ye, R. Quhe, Y. Pan, H. Zhang, H. Zhong, J. Shi and J. Lu, Many-body effect, carrier mobility, and device performance of hexagonal Arsenene and Antimonene, *Chem. Mater.*, 2017, **29**, 2191-2201.

30. B. Peng, H. Zhang, H. Shao, K. Xu, G. Ni, J. Li, H. Zhua and C. M. Soukoulis, Chemical intuition for high thermoelectric performance in monolayer black phosphorus, a-arsenene and aW-antimonene, *J. Mater. Chem. A*, 2018, **6**, 2018-2033.

31. D. Kecik, E. Durgun and S. Ciraci, Optical properties of single-layer and bilayer arsenene phases, *Phys. Rev. B*, 2016, **94**, 035423.

32. S. Nahas, A. Bajaj and S. Bhowmick, Polymorphs of two dimensional phosphorus and arsenic: An insight from evolutionary search, *Phys. Chem. Chem. Phys.*, 2017, **11**, 11282-11288.

33. J. Carrete, L. J. Gallego and N. Mingo, Structural complexity and phonon physics in 2D arsenenes, *J. Phys. Chem. Lett.*, 2017, **8**, 1375-1380.

34. P. Jamdagni, A. Kumar, A. Thakur, R. Pandey and P. K. Ahluwalia, Tunneling characterstics of Stone-Wales defects in monolayers of Sn and group-V elements, *J. Phys.: Condens. Matter*, 2017, **29**, 395501.

35. J. L. Zhang, S. Zhao, C. Han, Z. Wang, S. Zhong, S. Sun, R. Guo, X. Zhou, C. Gu, K. Yuan, Z. Li and W. Chen, Epitaxial growth of single layer blue phosphorus: A new phase of two-dimensional phosphorus, *Nano Lett.,* 2016 **16**, 4903-4908.

36. J. Guan, Z. Zhu and D. Tomanek, Semiconducting layered blue phosphorus: A computational study, *Phys. Rev. Lett.,* 2014, **112** 176802.

37. G. Kresse and J. Furthmuller, Efficient iterative schemes for *ab initio* total-energy calculations using a plane-wave basis set, *Phys. Rev. B*, 1996, **54**, 11169.

38. J. P. Perdew, K. Burke, and M. Ernzerhof, Generalized Gradient Approximation Made Simple, *Phys. Rev. Lett.*, 1996, **77**, 3865.

39. P. E. Blochl, Projector augmented-wave method, *Phys. Rev. B*, 1994, **50**, 17953.

40. S. Grimme, Semiempirical GGA-type density functional constructed with a long-range dispersion correction, *J. Comput. Chem.*, 2006, **27**, 1787.

41. J. Heyd, G. E. Scuseria and M. Ernzerh, Hybrid functionals based on screened Coulomb potential, *J. Chem. Phys.*, 2006, **118**, 8207.

42. J. Paier, M. Marsman, K. Hummer, G. Kresse, I. C. Gerber, and J. G. Angyan, Screened hybrid density functionals applied to solids, *J. Chem. Phys.*, 2006, **125**, 249901.

43. J. Bardeen and W. Shockley, Deformation potentials and mobilities in non-polar crystals, *Phys. Rev.*, 1950, **80**, 72-80.

44. H. Wang, Y. Pei, A. D. LaLonde and G J Snyder, Weak electron-phonon coupling contributing to high thermoelectric performance in n-type PbSe, *PANS*, 2012 **109**, 9705.

45. D. Schifer and C. S. Barrett, The crystal structure of arsenic at 4.2, 78 and $299^0$ K, *J. Appl. Crystallogr.*, 1969, **2**, 30.

46. H. Sahin, S. Cahangirov, M. Topsakal, E. Bekaroglu, E. Akturk, R. T. Senger, and S. Ciraci, Monolayer honeycomb structures of group-IV elements and III-V binary compounds: First-principles calculations, *Phys. Rev. B*, 2009, **80**, 155453.

47. A. Falin, Q. Cai, E. J.G. Santos, D. Scullion, D. Qian, R. Zhang, Z. Yang, S. Huang, K. Watanabe, T. Taniguchi, M. R. Barnett, Y. Chen, R. S. Ruoff and L. H. Li, Mechanical properties of atomically thin boron nitride and the role of interlayer interactions, *ACS Nano*, 2017, **8**, 15815.

48. S. Bertolazzi, J. Brivio and A. Kis, Stretching and Breaking of Ultrathin $MoS_2$, *ACS Nano*, 2011, **12**, 9703.

49. A. Kumar and P. K. Ahluwalia, Electronic Structure of Transition Metal Dichalcogenides Monolayer $1H-MX_2$ (M=Mo, W; X=S,Se,Te) from ab-initio Theory: New Direct Band Gap Semiconductors, *Eur. Phys. J. B*, 2012, **85**, 186.

50. P. Jamdagni, A. Kumar, A. Thakur, R. Pandey and P. K. Ahluwalia, Stability and electronic properties of SiGe-based 2D layered structures, *Mater. Res. Express*, 2015, **2**, 016301.

51. N. Petrone and J. Hone, Two-dimensional flexible nanoelectronics, *Nat. Commun.*, 2014, **5**, 5678.

52. V. Chakrapani, J. C. Angus, A. B. Anderson, S. D. Wolter, B. R. Stoner, G. U. Sumanasekera, Charge transfer equilibria between diamond and an aqueous oxygen electrochemical redox couple, *Science*, 2007, **318**, 1424.

53. B. Sa, Y. L. Li, J. Qi, R. Ahuja and Z. Sun, Strain engineering for phosphorene: The potential application as a photocatalys, *J. Phys. Chem. C.*, 2014, **118**, 26560-26568.

54. Y. Wang, M. Ye, M. Weng, J. Li, X. Zhang, H. Zhang, Y. Guo, Y. Pan, L. Xiao, J. Liu, F. Pan and J. Lu, Electrical contacts in monolayer arsenene devices, *ACS Appl. Mater. Interface,* 2017, **9**, 29273-29284

55. M. Hart, J. Chen, A. Michaelides, A. Sella, M. S. P. Shaffer and C. G. Salzmann, One-dimensional arsenic allotropes: polymerization of yellow arsenic inside single-wall carbon nanotubes, *Angew. Chem. Int. Ed.* 2018, **57**, 1-6

56. P. Vishnoi, M. Mazumder, S. K. Pati, and C. N. R. Rao, Arsenene nanosheets and nanodots, *New J. Chem.*, 2018, **42**, 14091-14095.